\documentclass[twocolumn,superscriptaddress,showpacs,aps,amsmath,amssymb,prb]{revtex4}
\usepackage{bm}
\usepackage{graphicx,color}
\usepackage{braket}

\begin{document}

\title{Lasing and transport in a multi-level double quantum dot system \\
coupled to a microwave oscillator}

\author{Christian Karlewski}
\email{christian.karlewski@kit.edu}
\affiliation{Institut f\"ur Theoretische Festk\"orperphysik, Karlsruhe Institute of Technology, D-76131 Karlsruhe, Germany}
\affiliation{Institute of Nanotechnology, Karlsruhe Institute of Technology, D-76344 Eggenstein-Leopoldshafen, Germany}
\author{Andreas Heimes}
\affiliation{Institut f\"ur Theoretische Festk\"orperphysik, Karlsruhe Institute of Technology, D-76131 Karlsruhe, Germany}
\author{Gerd Sch\"on}
\affiliation{Institut f\"ur Theoretische Festk\"orperphysik, Karlsruhe Institute of Technology, D-76131 Karlsruhe, Germany}
\affiliation{Institute of Nanotechnology, Karlsruhe Institute of Technology, D-76344 Eggenstein-Leopoldshafen, Germany}

\pacs{73.21.La, 42.50.Pq, 78.67.Hc, 85.35.Gv}

\date{\today}
\begin{abstract}
We study a system of two quantum dots, each with several discrete levels, which are coherently coupled to a microwave oscillator. 
They are attached to electronic leads and coupled to a phonon bath, both leading to inelastic processes.
For a simpler system with a single level in each dot it has been shown that a population inversion can be created by electron tunneling, which in a resonance situation leads to lasing-type properties of the oscillator.
In the multi-level system several resonance situations may arise, some of them relying on a sequence of tunneling processes which also involve non-resonant, inelastic transitions. 
The resulting photon number in the oscillator and the current-voltage characteristic are highly sensitive to these properties and accordingly can serve as a probe for microscopic details.
\end{abstract}

\maketitle

\section{Introduction}
In the last decade many concepts originally developed in the field of quantum optics, i.e., for atoms coupled to photons in optical cavities, could be demonstrated with electronic circuits where a single superconducting qubit is coupled to a microwave resonator \cite{451664a,nature06184,nature02831,nature02851,PhysRevA.69.062320}. 
These systems can be built with modern fabrication techniques and are interesting for applications because they are small enough to be integrated in chip structures.
Aided by the strong coupling which can be achieved, also the equivalent of single-atom lasing could be realized \cite{nature06141,PhysRevA.82.053802,NoriLasing}. The frequencies of these systems are in the GHz regime, accordingly they are sometimes called 'maser' instead of laser.  
Another setting, which is well controlled and understood but still shows unexplored physics, is a system of multi-level quantum dots coupled to an oscillator. Specifically we consider quantum dots made up of nanoscale semiconducting islands, which are attached to source and drain electrodes via tunnel junctions, with energy levels that can be shifted by capacitively coupled gates. Already a single quantum dot coupled to a microwave oscillator displays fascinating features like Kondo physics\cite{PhysRevLett.76.1715,PhysRevLett.107.256804} or the control of photon emission via transport properties 
\cite{PhysRevB.89.195127,PhysRevB.85.045446,PhysRevX.1.021009}.

\begin{figure}[tb]
 \includegraphics[origin=c,width=0.43\textwidth]{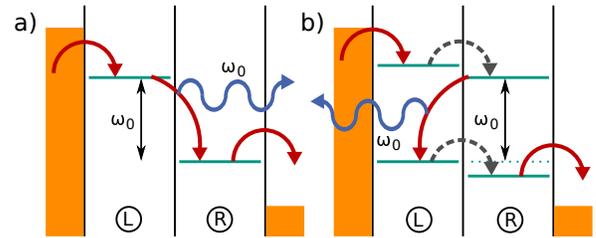}
 \caption{a) Sketch of the lasing-type situation with a double quantum dot and one level in each dot. Tunneling processes of electrons and photons emitted to the microwave oscillator are illustrated by solid and wavy arrows, respectively. b) With two levels in each dot a cascade of transitions and various resonance situations leading to lasing can be achieved. In these cases also inelastic transitions mediated by phonons, indicated by dashed arrows, are important.}
  \label{pSketch}
\end{figure}

Also double quantum dot (DQD) systems coupled to 
a microwave oscillator have been investigated already, both in theory \cite{PhysRevA.69.042302,1103.5051,JinMarthalerLasing,PhysRevB.71.205322,PhysRevLett.114.196802,1503.01597} as well as in experiments \cite{PhysRevLett.108.046807,PhysRevB.88.125312,285.full, PhysRevLett.113.036801,PhysRevLett.115.046802,ncomms4716,PhysRevB.89.165404}. 
By appropriately tuning the level structure of the DQD system with a single level each, 
as shown in Fig.~\ref{pSketch}a), a population inversion can be created, and in a resonance situation a lasing state can be produced\cite{PhysRevA.69.042302,1103.5051,JinMarthalerLasing}. This state is also strongly reflected in the transport properties. The theoretical predictions have been confirmed by recent experiments of Liu {\it et al.}\cite{285.full, PhysRevLett.113.036801}. A thorough description of the system also requires the analysis of the effects of further degrees of freedom leading to relaxation and decoherence. E.g., the effect of phonons has been studied. On the one hand, the additional bath leads to decoherence which deteriorates the performance of the device \cite{PhysRevB.71.205322}. On the other hand, the phonons enhance the transport through the DQD and hence the photon production \cite{PhysRevLett.114.196802,1503.01597,ncomms4716}.
 
With multiple levels in each dot of the DQD system electron transport may be achieved via a series of transitions, and different lasing situations may arise whenever the microwave oscillator is in resonance with one of the transitions. A special situation is illustrated in Fig.~\ref{pSketch}b), where four levels of the DQD are involved, with the resonant tunneling process against the total current direction. All other transitions inside the DQD are inelastic. They are allowed due to the coupling to phonons, which needs to be included in a description of the system.

In this article we analyze the properties of a voltage-biased multi-level DQD coherently coupled to a microwave oscillator with additional coupling to the phonons of the substrate. In Sec.~\ref{cMicro} we introduce the model Hamiltonian and present in detail the corresponding quantum master equation. In Sec.~\ref{cCascade} we discuss results with emphasis on the variety of lasing states and their signatures in the transport current. We use parameters close to those found in recent experiments\cite{PhysRevLett.113.036801}. We summarize  in Sec.~\ref{cConclusion}.

\section{The model Hamiltonian and master equation}
\label{cMicro}
We first review the quantum master equation describing the multi-level DQD system coupled to a microwave oscillator and electronic reservoirs in detail. In a second step we extend the theory to account for the coupling to phonons.

\subsection{Multi-level quantum dots coupled to a microwave cavity}
The DQD system with discrete energy levels $\varepsilon_{\alpha i}$ in the left and right ($\alpha=L, R$) dots coupled to a microwave oscillator and electronic leads is described by the Hamiltonian 
\begin{align}
 H=H_{\rm DQD}+H_{\rm osc}+H_{\rm DQD-osc}+H_{\rm C} + H_{\rm leads}.
\end{align}
The DQD Hamiltonian is
\begin{align}
 H_{\rm DQD}=&\sum_{\substack{\alpha=L,R\\ i}}\varepsilon_{\alpha i}\, n_{\alpha i}+\sum_{ij}\left(t_{ij}d_{Li}^\dagger d_{Rj}+\text{h.c.}\right)\notag\\
 &+\, U_1\sum_{\substack{\alpha=L,R\\ i\ne j}}n_{\alpha i}\, n_{\alpha j}+U_2\sum_{ij}n_{L i}\, n_{R j} \, ,
\end{align}
with $n_{\alpha i}= d_{\alpha i}^\dagger d_{\alpha i}$. Here $U_1$ and $U_2$ are the intra-dot and inter-dot Coulomb energy scales, respectively. We allow for tunneling between the two dots with amplitudes $t_{ij}$ which produces the current but also leads to hybridization of the dot states. 
The microwave oscillator is modeled having a single mode 
\begin{align}
 H_{\rm osc}=\omega_0a^\dagger a.
\end{align}
It is assumed to be coupled to the left dot with an energy-independent coupling constant g, i.e.
\begin{align}
 H_{\rm DQD-osc}=g (a+a^\dagger)\sum_i n_{Li}.
\end{align}
For later use we combine the three parts of the Hamiltonian which give rise to coherent time evolution to the system Hamiltonian $H_{\rm S}=H_{\rm DQD}+H_{\rm osc}+H_{\rm DQD-osc}$. 
We further assume that the Coulomb energy scales are large, which allows us to concentrate on states with at most one electron in the DQD system.

The couplings to the left and right lead are assumed to be weak and, for simplicity, equal for both sides, $\gamma_L=\gamma_R=\gamma$. Hence we have 
\begin{align}
 H_{\rm C}=\gamma\sum_{\substack{\alpha=L,R\\ ik}}\left(d^\dagger_{\alpha i}c_{\alpha k}+\text{h.c.}\right).
\end{align}

As usual, we assume the leads with Hamiltonian $H_{\rm leads}$ to be in equilibrium. The relevant expectation values are the Green's functions  given by
\begin{align}
 G^<_{\alpha k}(t)=&\, i\braket{c^\dagger_{\alpha k}(0)c_{\alpha k}(t)}=if(\epsilon_{\alpha k}+eV_\alpha)e^{-i(\epsilon_{\alpha k}+eV_\alpha)t}, \nonumber\\
 G^>_{\alpha k}(t)=&-i\braket{c_{\alpha k}(t)c^\dagger_{\alpha k}(0)}\notag\\
 =&-i\left[1-f(\epsilon_{\alpha k}+eV_\alpha)\right]e^{-i(\epsilon_{\alpha k}+eV_\alpha)t},
\end{align}
with Fermi-Dirac distribution function $f(\epsilon)$. 

To proceed we consider the reduced density matrix of the DQD and oscillator system, $\rho_{\rm S}(t)$, with the lead degrees of freedom traced out. In the limit of weak coupling to the electrodes and short correlation times inside the leads as compared to typical system time scales, a Liouville-von~Neumann equation in the Born-Markov approximation is sufficient,
\begin{align}
 \label{eLvNBM}
 \frac{\partial}{\partial t} \rho_{\rm S}(t)=\mathcal{L}\rho_{S}(t),
\end{align}
 with Liouvillian $\mathcal{L}=\mathcal{L}_{\rm S}+\mathcal{L}_{\rm C}$ given by
\begin{align}
 \label{eLiou}
 \mathcal{L}_{\rm S} \rho_{S}&(t)=\, i\left[\rho_{S}(t),H_{\rm S}\right],\\
 \nonumber\\
 \mathcal{L}_{\rm C} \rho_{S}&(t)=-\int \limits_{-\infty}^t\text{d}t'\braket{\left[H_{\rm C}(t),\left[H_{\rm C}(t'),\rho_{S}(t')\right]\right]}_{LR}.
\end{align}
The expectation value $\braket{\cdot}_{LR}=\text{Tr}\left\{\cdot\rho_L\rho_R\right\}$ is taken with respect to the equilibrium density matrices of the leads.

We are interested in the stationary solution of the quantum master equation
$\partial \rho_{st}/\partial t=0=\mathcal{L}\rho_{st}$. It can be written in terms of the Laplace transforms
\begin{align}
 G_{\alpha k}^<(\omega)= & \int \limits_{-\infty}^0\text{d}t \, G_{\alpha k}^<(t)e^{i\omega t+\eta t}=\frac{f(\epsilon_{\alpha k}+eV_\alpha)}{\omega-(\epsilon_{\alpha k}+eV_\alpha)-i\eta},\notag\\
 G_{\alpha k}^>(\omega)=&\int \limits_{-\infty}^0\text{d}t \, G_{\alpha k}^>(t)e^{i\omega t+\eta t}=-\frac{1-f(\epsilon_{\alpha k}+eV_\alpha)}{\omega+\epsilon_{\alpha k}+eV_\alpha-i\eta},
\end{align}
with the small parameter $\eta=0_+$ introduced for convergence. Next we perform a unitary transformation, which diagonalizes the Hamiltonian $H_{\rm S}$, with $V^\dagger H_{\rm S} V=\text{diag}(E_1,\ldots,E_N)$ and introduce the notation $\tilde{A}=V^\dagger AV$. With this the Lindblad operator $\mathcal{L}_{\rm C}$ becomes \cite{PhysRevB.90.104302,PhysRevB.85.174515} 
\begin{align}
 &V^\dagger\mathcal{L}_{\rm C}\rho_{st}V=i\gamma^2\sum_{\substack{\alpha=L,R \\ ijk}}\notag\\
 &\times \left\{\tilde{d}_{\alpha i}\left[G_{\alpha k}^<(\hat{\omega})*\tilde{d}_{\alpha j}^\dagger\right]\tilde{\rho}_{st}+\tilde{d}_{\alpha i}^\dagger\left[G_{\alpha k}^>(\hat{\omega})^\dagger*\tilde{d}_{\alpha j}\right]\tilde{\rho}_{st}\right.\notag\\
 &\quad+\tilde{d}_{\alpha i}\tilde{\rho}_{st}\left[G_{\alpha k}^>(\hat{\omega})*\tilde{d}_{\alpha j}^\dagger\right]+\tilde{d}_{\alpha i}^\dagger\tilde{\rho}_{st}\left[G_{\alpha k}^<(\hat{\omega})^\dagger*\tilde{d}_{\alpha i}\right]\notag\\
 &\quad-\left[G_{\alpha k}^>(\hat{\omega})^\dagger*\tilde{d}_{\alpha i}\right]\tilde{\rho}_{st}\tilde{d}_{\alpha j}^\dagger-\left[G_{\alpha k}^<(\hat{\omega})*\tilde{d}_{\alpha i}^\dagger\right]\tilde{\rho}_{st}\tilde{d}_{\alpha j}\notag\\
 &\left.-\tilde{\rho}_{st}\left[G_{\alpha k}^<(\hat{\omega})^\dagger*\tilde{d}_{\alpha i}\right]\tilde{d}_{\alpha j}^\dagger-\tilde{\rho}_{st}\left[G_{\alpha k}^>(\hat{\omega})*\tilde{d}_{\alpha i}^\dagger\right]\tilde{d}_{\alpha j}\right\}
\end{align}
The product denoted by the symbol $*$ represents the direct or Hadamard product between two matrices, e.g.,
\begin{align}
 \left(G_{\alpha k}^>(\hat{\omega})^\dagger*\tilde{d}_{\alpha j}\right)_{nm}
=\left(G_{\alpha k}^{>\phantom{\dagger}}(\hat{\omega})\right)^*_{mn}(\tilde{d}_{\alpha j})_{nm}.
\end{align}
In addition we introduced the matrix $\hat{\omega}$ with  elements $\hat{\omega}_{nm}=E_n-E_m$.

Decay processes in the oscillator, characterized by the decay rate $\kappa$, are described as usual by the Lindblad operator
\begin{align}
 \mathcal{L}_{\kappa}\rho_{st}=\frac{\kappa}{2}\left(2a\rho_{st}a^\dagger-\left[a^\dagger a,\rho_{st}\right]_+\right),
\end{align}
where $[\cdot,\cdot]_+$ denotes the anti-commutator.

Physical observables of interest are the current through the system and the number of photons in the oscillator, as well as their statistical properties.
In order to evaluate the current we consider the time evolution of the electron number in the left lead $N_L(t)=\sum_{k}c_{Lk}^{\dagger}(t)c_{Lk}(t)$. This operator commutes with all parts of the Hamiltonian except for the coupling term $H_{\rm C}$. We thus get for the current through the left tunnel junction
\begin{align}
\braket{I_L}= & \left.e\frac{d}{dt}\langle N_L(t)\rangle\right|_{t\rightarrow\infty}=\left.ie\langle[H(t),N_L(t)]\rangle\right|_{t\rightarrow\infty}\notag\\
 = &\, 4 e t^2\sum_{kij}\text{Tr}\left\{\text{Im}\left[-\tilde{d}_{Li}\left(G^<_{Lk}(\hat{\omega})*\tilde{d}_{Li}^\dagger\right)\tilde{\rho}_{st}\right.\right.\notag\\
 &\left.\left.+\, \tilde{d}_{Li}^\dagger\left(\tilde{d}_{Lj}*G_{Lk}^>(\hat{\omega})^\dagger\right)\tilde{\rho}_{st}\right]\right\},
\end{align}

The state of the oscillator is characterized by the number of photons and the Fano factor \cite{RevModPhys.29.74} given by
\begin{align}
 \braket{N_{Ph}}=&\text{Tr}\left\{a^\dagger a \, \rho_{st}\right\}\\
 F_a=&\left(\braket{N_{Ph}^2}-\braket{N_{Ph}}^2\right)/\braket{N_{Ph}} .
\end{align}
In the lasing state the number of photons is strongly enhanced with a characteristic narrow peak. Additionally, the Fano factor is an indicator for lasing. In the non-lasing regime the Fano factor is approximately $F_a\approx 1+\braket{N_{Ph}}$. In the lasing regime the Fano factor should drop strongly. For ideal lasing one has $F_a=1$, indicative of a Poissonian process, whereas in real systems we may observe deviations, even smaller values, $F_a<1$, corresponding to a sub-Poissonian distribution of the radiation field \cite{1103.5051}.

\subsection{Coupling to phonons}
\label{cMicroPhon}
Electron transport through the DQD system may be influenced by the coupling to phonons of the semiconducting bulk material \cite{PhysRevLett.114.196802,PhysRevLett.113.036801,ncomms4716}.
We assume that electrons in the DQD couple to bulk phonon modes via a piezoelectric interaction 
\cite{1-s2.0-S0370157304005496-main}. The Hamiltonian  is
\begin{align}
 H_{\rm el-ph}=g_{\rm el-ph}\sum_{\alpha ik}n_{\alpha i}\varphi_{\alpha k} 
\end{align}
with phonon operators  $\varphi_{\alpha k}=b_{\alpha k}+b_{\alpha -k}^\dagger$ and the coupling constant $g_{\rm el-ph}$, which is considered to be energy-independent. 
The phonon-bath is assumed to remain in equilibrium such that the phonon Green's functions are given by
\begin{align}
 D_{\alpha k}^<(t)=&\, i\braket{\varphi_{\alpha k}(0)\varphi_{\alpha k}(t)}\notag\\
 = &\, i n_B(\omega_{\alpha k})e^{-i\omega_{\alpha k}t}+i[1+n_B(\omega_{\alpha k})]e^{i\omega_{\alpha k}t},\notag\\
 D_{\alpha k}^>(t)=&-i\braket{\varphi_{\alpha k}(t)\varphi_{\alpha k}(0)}\notag\\
 =&-in_B(\omega_{\alpha k})e^{i\omega_{\alpha k}t}-i[1+n_B(\omega_{\alpha k})]e^{-i\omega_{\alpha k}t}\, ,
\end{align}
with the Bose-Einstein distribution $n_B(\omega)$. The respective Laplace transforms are
\begin{align}
 D_{\alpha k}^<(\omega)=&\frac{n_B(\omega_{\alpha k})}{\omega-\omega_{\alpha k}-i\eta}+\frac{1+n_B(\omega_{\alpha k})}{\omega+\omega_{\alpha k}-i\eta},\notag\\
 D_{\alpha k}^>(\omega)=&-\frac{n_B(\omega_{\alpha k})}{\omega+\omega_{\alpha k}-i\eta}-\frac{1+n_B(\omega_{\alpha k})}{\omega-\omega_{\alpha k}-i\eta}.
\end{align}
In the Born-Markov approximation of the quantum master equation the phonons are implemented by an additional Liouville operator
\begin{align}
 &V^\dagger\mathcal{L}_{\rm el-ph}\rho_{st}V=ig_{\rm el-ph}^2\sum_{\alpha ijk}\notag\\
 & \times \left\{\tilde{n}_{\alpha i}\left[\vphantom{\tilde{d}_{\alpha j}^\dagger}D_{\alpha k}^<(\hat{\omega})*\tilde{n}_{\alpha j}\right]\tilde{\rho}_{st}\right.
-\tilde{n}_{\alpha i}\tilde{\rho}_{st}\left[\vphantom{\tilde{d}_{\alpha j}^\dagger}D_{\alpha k}^>(\hat{\omega})*\tilde{n}_{\alpha j}\right]\notag\\
&-\left[\vphantom{\tilde{d}_{\alpha j}^\dagger}D_{\alpha k}^<(\hat{\omega})*\tilde{n}_{\alpha i}\right]\tilde{\rho}_{st}\tilde{n}_{\alpha j}
- \left.\tilde{\rho}_{st}\left[D_{\alpha k}^>(\hat{\omega})*\tilde{n}_{\alpha i}\vphantom{\tilde{d}_{\alpha j}^\dagger}\right]\tilde{n}_{\alpha k}\right\},
\end{align}
with 
the sum over k to be evaluated with the effective phonon density of states  $F(\omega)$.

In summary, we look for stationary solutions of the quantum master equation, $\mathcal{L}\rho_{st}=0$, with total Liouvillian of the DQD and oscillator with dissipation due to the electron tunneling, phonon coupling and decay of the oscillator given by
\begin{align}
\mathcal{L}\rho_{st}=\left(\mathcal{L}_{\rm S}+\mathcal{L}_{\rm C}+\mathcal{L}_{\rm el-ph}+\mathcal{L}_{\kappa}\right)\rho_{st}. \label{eFME}
\end{align}
In the following section we will analyze this equation numerically with the choice of parameters inspired by recent experiments.

\section{Lasing in the multi-level system}
\label{cCascade}
\subsection{Results for various resonance situations}
Lasing in a double quantum dot--oscillator system has been recently observed in experiments of 
Liu \textit{et al.}\cite{PhysRevLett.113.036801}. We will use the parameter found in these experiments as starting point for our numerical investigation with small adjustments to better emphasize our findings.
In the following example we consider an oscillator with frequency $f_0 = 7.8 \,\mathrm{GHz} =\omega_0/2\pi$, i.e., $\omega_0= 0.032\,\mathrm{meV}$ as in these experiments. This energy serves as reference scale for all other parameters. 
The applied voltages $V_\alpha$, leading to electron transport from the left to the right lead, are assumed
large enough that several levels of the DQD system lie in the window $eV_L > \epsilon_{Li},\epsilon_{Ri} > eV_R$.
The left dot is assumed to be coupled to the oscillator with strength $g=0.02 \,\omega_0$.
This is an order of magnitude stronger than found in the experiment, but we choose this value for a better display of the relevant properties.
For the same reason, we consider a very high-quality microwave oscillator with $\kappa=10^{-6} \, \omega_0$ which corresponds to  a quality factor $Q=2\cdot10^{6}$. Such a quality has been achieved in purely superconducting systems \cite{Megrant2012}, while in semiconductor-superconductor heterostructures only values reaching $Q\approx10^4$ have been reported \cite{oe-16-8-5199}.

To account for unavoidable heating effects we choose the temperature to be $T=100\,\mathrm{mK}=0.27 \, \omega_0$, i.e., a factor 10 higher than the base temperature in the experiment. 
In the experiments Liu \textit{et al.}\cite{PhysRevLett.113.036801} also the effective phonon density of states $F_{\alpha}(\omega)=\sum_{\alpha k}\delta(\omega-\omega_{\alpha k})$ has been extracted. It is given approximately by 
\begin{align}
 F(\omega)=\frac{(\omega/\Omega_0)^2}{[\alpha_1+(\omega/\Omega_0)^2]^{\alpha_2/2+1}}
\end{align}
with parameters $\alpha_1=0.02$, $\alpha_2=1.4$, and $\Omega_0=0.4\,\mathrm{meV}$.

For definiteness we focus on the situation illustrated in Fig.~\ref{pSketch}b) with two levels in each dot
lying in the window between the applied potentials. The two levels in the left dot are separated by the energy $\varepsilon_{L2}-\varepsilon_{L1}=\Delta_L\omega_0$ and similar in the right dot, $\varepsilon_{R2}-\varepsilon_{R1}=\Delta_R\omega_0$. 
In general, the energy splittings  differ from the oscillator frequency ($\Delta_{L/R} \ne 1$), and they differ from each other ($\Delta_L\neq\Delta_R$). We denote the relative shift of the two lower energy levels by  $\varepsilon_{L1}-\varepsilon_{R1}=\varepsilon$. This parameter can be varied by applied gate voltages.
For simplicity, the tunneling amplitude between the left and right dot levels are assumed to be the same for all levels, $t_{ij}=t$ with values varied in the range $0.01\omega_0$ to $0.16\omega_0$ in the following. Furthermore, we fix the electron tunneling rate to the leads $\Gamma=2\pi N_0(E_f)\gamma^2$, with electron DOS at the Fermi edge $N_0(E_f)$, at the value $\Gamma=10^{-6}\omega_0$. It appears for small $\Gamma$ predominantly as an overall scaling factor of the current.

\begin{figure}[t]
\includegraphics[origin=c,width=0.48\textwidth]{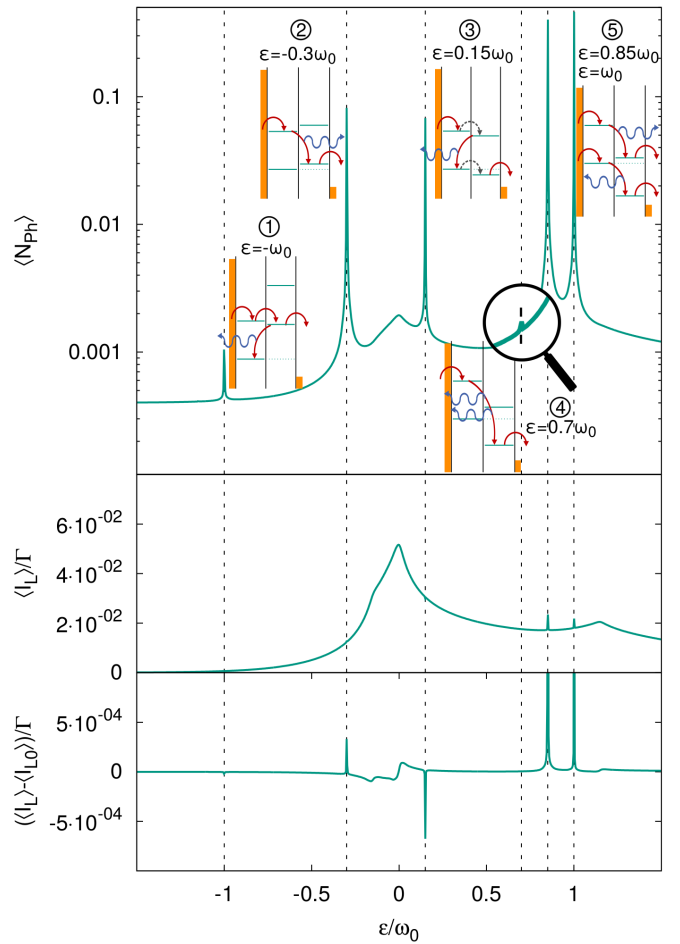}
\caption{The upper figure shows the photon expectation value $\braket{N_{Ph}}$ vs. the detuning $\varepsilon$ with $\Delta_L=1.3$, $\Delta_R=1.15$. The insets show sketches of the different energy level arrangements. The continues arrows represent electron hopping, the dashed arrows phonon induced hopping and the curly arrows photon emission. The vertical dashed lines mark various resonance situations.  The parameters are $t=0.01\omega_0$ and $g_{\rm el-ph}=0.01\omega_0$. The middle figure shows the current through the left lead $\braket{I_L}/\Gamma$ vs. the detuning $\varepsilon$. The lower part shows the difference between the current through the left lead with ($g=0.01\omega_0$) and without ($g=0$) coupling to the oscillator. The current is enhanced or decreased in the resonance situations.}
 \label{pPhot_overview}
\end{figure}

In Fig.~\ref{pPhot_overview} results are plotted for varying values of the relative shift $\varepsilon$ of the levels in the two dots. The upper part of the figure shows the photon number $\braket{N_{Ph}}$. It peaks whenever one of the downwards electronic transitions in the DQD is in resonance with the microwave oscillator. In combination with the increased population of the higher level due to the transport current a lasing-type situation arises (see below). The sketches show the arrangement of energy levels and the electronic transitions responsible for the resonant excitations of photons. 

The two peaks on the right (corresponding to the sketch labeled by (5)) occur at $\varepsilon=\omega_0$ and $\varepsilon=(1-\Delta_L+\Delta_R)\omega_0$, when the transitions between the two lower or the two higher energy levels of the dots are in resonance with the microwave oscillator. This is a straightforward extension of what had been found for a double-dot system with a single level in each dot \cite{1103.5051}. The same is true for the peak 
at $\varepsilon=(1-\Delta_L)\omega_0$ labeled by (2), where the resonant transition is between the upper level of the left dot and the lower one in the right dot.

A completely different situation is found for the peak at  $\varepsilon=(\Delta_R-1)\omega_0$ (labeled by (3))
Here, a cascade of processes is involved: (i) an incoherent transition, associated with the excitation of a phonon, from the higher left dot level to the higher right dot level, (ii) a transition to the lower left dot level in resonance with the microwave oscillator, (iii) another incoherent transition, again associated with the emission of a phonon, to the lower right dot level. Note that in this process the resonant electron tunneling is against the main current direction.

The small peak at the left (labeled by (1)) arises in the resonance situation 
when the transition from the lower level of the right dot to the lower level of the left dot is in resonance with 
the microwave oscillator. The peak is small since for the cycle to continue the electron has to escape to the right lead in a higher order or thermally activated process via one of the right dot levels. 

Another peak at $\varepsilon=(2-\Delta_L)\omega_0$ (labeled by (4)) is due to a two-photon resonance process ($\epsilon_{L2} -\epsilon_{R1} = 2 \omega_0$). It is very small for the chosen coupling strength between the DQD and the microwave oscillator but becomes more pronounced with increasing coupling strength.

The middle panel of Fig.~\ref{pPhot_overview} shows the current as a function of $\epsilon$. With our chosen paramters the current $\braket{I_L}$ is in the order of $1\,\mathrm{fA}$. The dominant feature in the center is a resonant tunneling peak when the lower two levels in the DQD coincide. There is another peak when the upper two levels are aligned. However, for the choice of parameters with $\Delta_L > \Delta_R$ in this situation the lower left level lies below the levels in the right dot, which means that once it is occupied the Coulomb interaction blocks further transport until a higher order or thermally activated process allows an escape to the right lead. This Coulomb blocking reduces the weight of the second resonance tunneling peak. It is in fact the reason why the current is small for all $\epsilon < 0$ compared to the current for $\epsilon > 0$. At the far right we see another resonance peak when the lower level in the left dot is aligned with the higher level in the right dot. For the chosen parameters the linewidth of these resonance tunnel peaks is determined by the temperature. In addition we see  much sharper peaks in the lasing-type situations when one of the electronic transitions is in resonance with the microwave oscillator. Some of them are two small or too close to the resonance tunneling peaks to be observable at the resolution of the plot. The more prominent ones at $\epsilon/\omega_0=0.85$ and $1$ arise in the two lasing situations labeled by (5) in the upper panel.

Further lasing resonances do show up in the current. In order to display them we plot in the lower of the three panels in Fig.~\ref{pPhot_overview} with higher resolution the difference of the current between the situation with coupling to the oscillator and without. This plot also displays the novel property that lasing resonance peaks in the photon number associated with backward tunneling processes (labeled (3) and (1)) lead to a sharp dip in the current.

\subsection{Dependence on further parameters}
\begin{figure}[t]
\includegraphics[origin=c,width=0.48\textwidth]{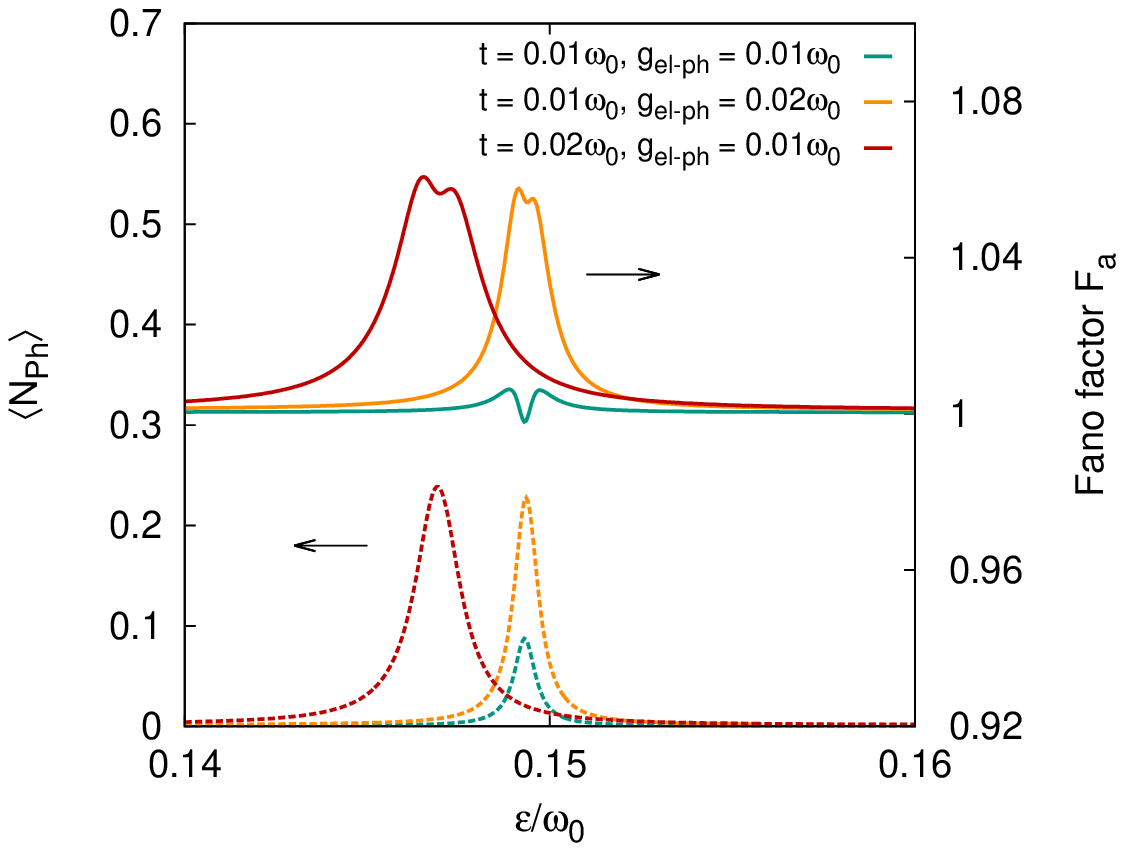}
\caption{Photon expectation value $\braket{N_{Ph}}$ (dashed lines) and Fano factor $F_a$ (continues lines) vs.  detuning $\varepsilon$ for $\Delta_L=1.3$, $\Delta_R=1.15$ and different phonon couplings strengths $g_{\rm el-ph}$ or tunneling strengths $t$.}
 \label{pFano}
\end{figure}

To test whether and to which extent the observed photon peaks are associated with lasing we analyze the Fano factor $F_a$. In Fig.~\ref{pFano} we show the results in the vicinity of the resonance labeled (3), but similar results are found also for the other lasing peaks. The figure displays the increase of the Fano factor in the vicinity of all peaks. For weak tunneling strength $t=0.01\omega_0$ and phonon coupling strength $g_{\rm el-ph}=0.01\omega_0$, the Fano factor shows a clear dip and even drops below 1. This is a signature of lasing, although the total number of photons remains small. With increasing phonon coupling or tunneling strength the number of photon rises, but the dip is less pronounced. The small asymmetry of the peaks and the Fano factor is related by the increasing conductivity of the dot system with decreasing $\varepsilon$ (cf. Fig. \ref{pPhot_overview}).

\begin{figure}[hbt]
\includegraphics[origin=c,width=0.48\textwidth]{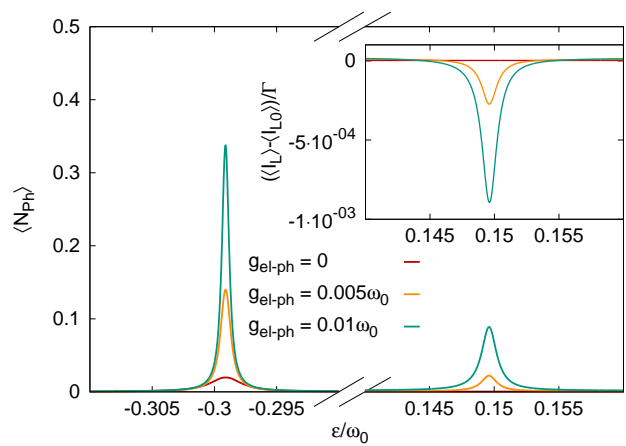}
\caption{Photon number $\braket{N_{Ph}}$ vs.  detuning $\varepsilon$ for $\Delta_L=1.3$, $\Delta_R=1.15$ and  different phonon coupling strengths $g_{\rm el-ph}$. The tunneling strength is $t=0.01\omega_0$. The inset shows the difference $(\braket{I_L}-\braket{I_{L0}})/\Gamma$ between the currents with ($g=0.01\omega_0$) and without ($g=0$) coupling to the oscillator  vs. the detuning $\varepsilon$.}
 \label{pPhot_Phonons}
\end{figure}
The processes involving a series of transitions, incl.\ inleastic ones, and the resulting lasing-type situations strongly rely on the coupling to the phonons. We illustrate this in Fig.~\ref{pPhot_Phonons} for the two processes labeled (2) at $\varepsilon=(1-\Delta_L)\omega_0$  and (3) at $\varepsilon=(\Delta_R-1)\omega_0$ by varying the strength of phonon coupling. The peak in the photon number, arising from a backward tunneling process which relies on the population of the upper right level, disappears in the absence of phonons and grows with increasing phonon coupling strength. 
The effect is also visible in the current as shown in the inset of the figure.
Also in the process (2) the transition through the DQD system is strongly enhanced by the phonons, 
which leads to more pronounced photon peaks. 
\begin{figure}[t]
\includegraphics[origin=c,width=0.48\textwidth]{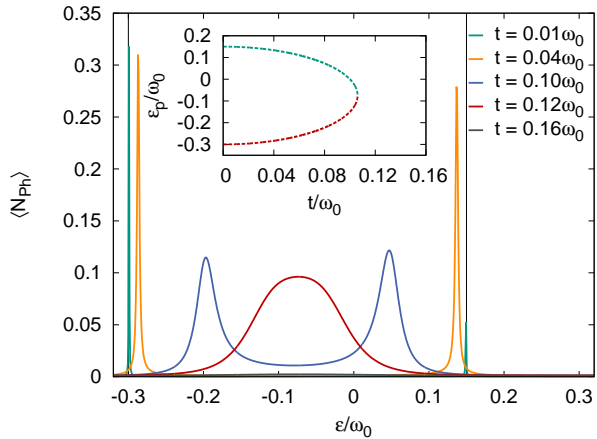}
\caption{Photon number $\braket{N_{Ph}}$ vs. the detuning $\varepsilon$ with $\Delta_L=1.3$, $\Delta_R=1.15$ and for different tunneling strengths $t$. The phonon coupling strength is $g_{\rm el-ph}=0.01\omega_0$. The inset shows the two ideal solutions for the peak position $\varepsilon_p$ vs. the tunneling strength $t$.}
 \label{pPhot_Tunneling}
\end{figure}

In Fig.~\ref{pPhot_Tunneling} we illustrate the dependence of the lasing peaks on the tunneling strength $t$. Such a dependence was also checked in the experiments with single-level double quantum dots by Stockklauser \textit{et al.} \cite{PhysRevLett.115.046802}. They found that with stronger tunneling strength the hybridization of the levels leads to a merging of the resonance peaks. In our system, for weak tunneling the lasing peaks are very sharp. With increasing tunneling strength the transport current through the DQD and the population inversion increases which leads to more photons, but eventually also the hybridization of the dot levels becomes stronger. This leads to broader and weaker peaks, and the peaks are shifted due to the hybridization. Their positions follow from the condition  
\begin{align}
1=&\,\tfrac{\Delta_L+\Delta_R}{2}-\sqrt{(\tfrac{t}{\omega_0})^2+\left(\tfrac{\varepsilon_p}{2\omega_0}\right)^2}\notag\\
&-\sqrt{(\tfrac{t}{\omega_0})^2+\left(\tfrac{\Delta_L-\Delta_R+\varepsilon_p/\omega_0}{2}\right)^2}.
\end{align}
This equation has two solutions for not too strong tunneling $t$ (as illustrated in the inset of Fig.~\ref{pPhot_Tunneling}) which determine the positions of the two lasing peaks. When the tunneling strength is  stronger the peaks merge and it is not possible to satisfy the separate resonance conditions anymore. In this case we still observe an enhanced photon number, but the process is no longer associated with lasing. A physical argument for the disappearing of the lasing peaks may be as follows: For strong hybridization the distinction between left and right levels is lost. This is, however, crucial for the creation of a population inversion by the imposed current.

\section{Summary}
\label{cConclusion}
Electron transport through a double quantum dot system coupled to a microwave oscillator may lead to a lasing state (strictly speaking a 'masing' state) with a narrow resonance peak in the photon number and a tendency to a Poisson distribution\cite{PhysRevA.69.042302,1103.5051}. It is interesting that this property of the photon field is reflected also in a peak in the transport current, which provides an alternative to study the lasing state. In the present paper we extended earlier work devoted to dots with one level each \cite{1103.5051} to dots with multiple levels. As expected we see several peaks of the previously known type but also qualitatively different ones. In particular we observe that in a cascade of transitions a backward tunneling process leads to a lasing peak in the photon number which is associated with a dip in the transport current. These multi-level  processes involve inelastic transitions, which may occur in the presence of phonons.
We therefore analyzed the multilevel DQD -- microwave oscillator system coupled to piezoelectric acoustic phonons. We used realistic parameters, taken from the experiments of Liu \textit{et al.}\cite{PhysRevLett.113.036801} including the phonon spectrum as determined in this reference. The lasing peaks and Fano factor are sensitive to the coupling strength and spectrum and may be a useful tool to analyze phonon properties in more detail.

Theoretically we analyzed the system in the frame of a quantum master equation, extending earlier work to multiple levels and a detailed analysis of the coupling to phonons. The description covers various quantum properties, such as the lasing, but we also see the effect of two-photon transitions. Further physical properties of the system, e.g., spin-degrees of freedom leading to spin-blockade effects \cite{nature11559} or spin-photon coupling \cite{Science-2015-Viennot-408-11}, could be implemented in this framework. 

\section{Acknowledgments}
We thank J. Jin, P.-Q. Jin, V. F. Maisi, M. Marthaler, I. Schwenk and A. Shnirman for stimulating discussions.


\begin{thebibliography}{34}
\expandafter\ifx\csname natexlab\endcsname\relax\def\natexlab#1{#1}\fi
\expandafter\ifx\csname bibnamefont\endcsname\relax
  \def\bibnamefont#1{#1}\fi
\expandafter\ifx\csname bibfnamefont\endcsname\relax
  \def\bibfnamefont#1{#1}\fi
\expandafter\ifx\csname citenamefont\endcsname\relax
  \def\citenamefont#1{#1}\fi
\expandafter\ifx\csname url\endcsname\relax
  \def\url#1{\texttt{#1}}\fi
\expandafter\ifx\csname urlprefix\endcsname\relax\def\urlprefix{URL }\fi
\providecommand{\bibinfo}[2]{#2}
\providecommand{\eprint}[2][]{\url{#2}}

\bibitem[{\citenamefont{Schoelkopf and Girvin}(2008)}]{451664a}
\bibinfo{author}{\bibfnamefont{R.~J.} \bibnamefont{Schoelkopf}}
  \bibnamefont{and} \bibinfo{author}{\bibfnamefont{S.~M.}
  \bibnamefont{Girvin}}, \bibinfo{journal}{Nature}
  \textbf{\bibinfo{volume}{451}}, \bibinfo{pages}{664} (\bibinfo{year}{2008}).

\bibitem[{\citenamefont{Majer et~al.}(2007)\citenamefont{Majer, Chow, Gambetta,
  Koch, Johnson, Schreier, Frunzio, Schuster, Houck, Wallraff
  et~al.}}]{nature06184}
\bibinfo{author}{\bibfnamefont{J.}~\bibnamefont{Majer}},
  \bibinfo{author}{\bibfnamefont{J.~M.} \bibnamefont{Chow}},
  \bibinfo{author}{\bibfnamefont{J.~M.} \bibnamefont{Gambetta}},
  \bibinfo{author}{\bibfnamefont{J.}~\bibnamefont{Koch}},
  \bibinfo{author}{\bibfnamefont{B.~R.} \bibnamefont{Johnson}},
  \bibinfo{author}{\bibfnamefont{J.~A.} \bibnamefont{Schreier}},
  \bibinfo{author}{\bibfnamefont{L.}~\bibnamefont{Frunzio}},
  \bibinfo{author}{\bibfnamefont{D.~I.} \bibnamefont{Schuster}},
  \bibinfo{author}{\bibfnamefont{A.~A.} \bibnamefont{Houck}},
  \bibinfo{author}{\bibfnamefont{A.} \bibnamefont{Wallraff}},
  \bibinfo{author}{\bibfnamefont{A.} \bibnamefont{Blais}},
  \bibinfo{author}{\bibfnamefont{M.~H.} \bibnamefont{Devoret}},
  \bibinfo{author}{\bibfnamefont{S.~M.} \bibnamefont{Girvin}},
  \bibinfo{author}{\bibfnamefont{R.~J.}~\bibnamefont{Schoelkopf}}, \bibinfo{journal}{Nature}
  \textbf{\bibinfo{volume}{449}}, \bibinfo{pages}{443} (\bibinfo{year}{2007}).

\bibitem[{\citenamefont{Chiorescu et~al.}(2004)\citenamefont{Chiorescu, Bertet,
  Semba, Nakamura, Harmans, and Mooij}}]{nature02831}
\bibinfo{author}{\bibfnamefont{I.}~\bibnamefont{Chiorescu}},
  \bibinfo{author}{\bibfnamefont{P.}~\bibnamefont{Bertet}},
  \bibinfo{author}{\bibfnamefont{K.}~\bibnamefont{Semba}},
  \bibinfo{author}{\bibfnamefont{Y.}~\bibnamefont{Nakamura}},
  \bibinfo{author}{\bibfnamefont{C.~J. P.~M.} \bibnamefont{Harmans}},
  \bibnamefont{and} \bibinfo{author}{\bibfnamefont{J.~E.} \bibnamefont{Mooij}},
  \bibinfo{journal}{Nature} \textbf{\bibinfo{volume}{431}},
  \bibinfo{pages}{159} (\bibinfo{year}{2004}).

\bibitem[{\citenamefont{Wallraff et~al.}(2004)\citenamefont{Wallraff, Schuster,
  Blais, Frunzio, Huang, Majer, Kumar, Girvin, and Schoelkopf}}]{nature02851}
\bibinfo{author}{\bibfnamefont{A.}~\bibnamefont{Wallraff}},
  \bibinfo{author}{\bibfnamefont{D.~I.} \bibnamefont{Schuster}},
  \bibinfo{author}{\bibfnamefont{A.}~\bibnamefont{Blais}},
  \bibinfo{author}{\bibfnamefont{L.}~\bibnamefont{Frunzio}},
  \bibinfo{author}{\bibfnamefont{R.-S.} \bibnamefont{Huang}},
  \bibinfo{author}{\bibfnamefont{J.}~\bibnamefont{Majer}},
  \bibinfo{author}{\bibfnamefont{S.}~\bibnamefont{Kumar}},
  \bibinfo{author}{\bibfnamefont{S.~M.} \bibnamefont{Girvin}},
  \bibnamefont{and} \bibinfo{author}{\bibfnamefont{R.~J.}
  \bibnamefont{Schoelkopf}}, \bibinfo{journal}{Nature}
  \textbf{\bibinfo{volume}{431}}, \bibinfo{pages}{162} (\bibinfo{year}{2004}).

\bibitem[{\citenamefont{Blais et~al.}(2004)\citenamefont{Blais, Huang,
  Wallraff, Girvin, and Schoelkopf}}]{PhysRevA.69.062320}
\bibinfo{author}{\bibfnamefont{A.}~\bibnamefont{Blais}},
  \bibinfo{author}{\bibfnamefont{R.-S.} \bibnamefont{Huang}},
  \bibinfo{author}{\bibfnamefont{A.}~\bibnamefont{Wallraff}},
  \bibinfo{author}{\bibfnamefont{S.~M.}~\bibnamefont{Girvin}}, \bibnamefont{and}
  \bibinfo{author}{\bibfnamefont{R.~J.} \bibnamefont{Schoelkopf}},
  \bibinfo{journal}{Phys. Rev. A} \textbf{\bibinfo{volume}{69}},
  \bibinfo{pages}{062320} (\bibinfo{year}{2004}).

\bibitem[{\citenamefont{Astafiev et~al.}(2007)\citenamefont{Astafiev, Inomata,
  Niskanen, Yamamoto, Pashkin, Nakamura, and Tsai}}]{nature06141}
\bibinfo{author}{\bibfnamefont{O.}~\bibnamefont{Astafiev}},
  \bibinfo{author}{\bibfnamefont{K.}~\bibnamefont{Inomata}},
  \bibinfo{author}{\bibfnamefont{A.~O.} \bibnamefont{Niskanen}},
  \bibinfo{author}{\bibfnamefont{T.}~\bibnamefont{Yamamoto}},
  \bibinfo{author}{\bibfnamefont{Y.~A.} \bibnamefont{Pashkin}},
  \bibinfo{author}{\bibfnamefont{Y.}~\bibnamefont{Nakamura}}, \bibnamefont{and}
  \bibinfo{author}{\bibfnamefont{J.~S.} \bibnamefont{Tsai}},
  \bibinfo{journal}{Nature} \textbf{\bibinfo{volume}{449}},
  \bibinfo{pages}{588} (\bibinfo{year}{2007}).

\bibitem[{\citenamefont{Andr\'e et~al.}(2010)\citenamefont{Andr\'e, Jin,
  Brosco, Cole, Romito, Shnirman, and Sch\"on}}]{PhysRevA.82.053802}
\bibinfo{author}{\bibfnamefont{S.}~\bibnamefont{Andr\'e}},
  \bibinfo{author}{\bibfnamefont{P.-Q.} \bibnamefont{Jin}},
  \bibinfo{author}{\bibfnamefont{V.}~\bibnamefont{Brosco}},
  \bibinfo{author}{\bibfnamefont{J.~H.} \bibnamefont{Cole}},
  \bibinfo{author}{\bibfnamefont{A.}~\bibnamefont{Romito}},
  \bibinfo{author}{\bibfnamefont{A.}~\bibnamefont{Shnirman}}, \bibnamefont{and}
  \bibinfo{author}{\bibfnamefont{G.}~\bibnamefont{Sch\"on}},
  \bibinfo{journal}{Phys. Rev. A} \textbf{\bibinfo{volume}{82}},
  \bibinfo{pages}{053802} (\bibinfo{year}{2010}).

\bibitem[{\citenamefont{Liu et~al.}(2004)\citenamefont{Liu, Wei, and
  Nori}}]{NoriLasing}
\bibinfo{author}{\bibfnamefont{Y.-X.} \bibnamefont{Liu}},
  \bibinfo{author}{\bibfnamefont{L.~F.} \bibnamefont{Wei}}, \bibnamefont{and}
  \bibinfo{author}{\bibfnamefont{F.}~\bibnamefont{Nori}},
  \bibinfo{journal}{Europhys. Lett.} \textbf{\bibinfo{volume}{67}},
  \bibinfo{pages}{941} (\bibinfo{year}{2004}).

\bibitem[{\citenamefont{K\"onig et~al.}(1996)\citenamefont{K\"onig, Schoeller,
  and Sch\"on}}]{PhysRevLett.76.1715}
\bibinfo{author}{\bibfnamefont{J.}~\bibnamefont{K\"onig}},
  \bibinfo{author}{\bibfnamefont{H.}~\bibnamefont{Schoeller}},
  \bibnamefont{and} \bibinfo{author}{\bibfnamefont{G.}~\bibnamefont{Sch\"on}},
  \bibinfo{journal}{Phys. Rev. Lett.} \textbf{\bibinfo{volume}{76}},
  \bibinfo{pages}{1715} (\bibinfo{year}{1996}).

\bibitem[{\citenamefont{Delbecq et~al.}(2011)\citenamefont{Delbecq, Schmitt,
  Parmentier, Roch, Viennot, F\'eve, Huard, Mora, Cottet, and
  Kontos}}]{PhysRevLett.107.256804}
\bibinfo{author}{\bibfnamefont{M.~R.} \bibnamefont{Delbecq}},
  \bibinfo{author}{\bibfnamefont{V.}~\bibnamefont{Schmitt}},
  \bibinfo{author}{\bibfnamefont{F.~D.} \bibnamefont{Parmentier}},
  \bibinfo{author}{\bibfnamefont{N.}~\bibnamefont{Roch}},
  \bibinfo{author}{\bibfnamefont{J.~J.} \bibnamefont{Viennot}},
  \bibinfo{author}{\bibfnamefont{G.}~\bibnamefont{F\'eve}},
  \bibinfo{author}{\bibfnamefont{B.}~\bibnamefont{Huard}},
  \bibinfo{author}{\bibfnamefont{C.}~\bibnamefont{Mora}},
  \bibinfo{author}{\bibfnamefont{A.}~\bibnamefont{Cottet}}, \bibnamefont{and}
  \bibinfo{author}{\bibfnamefont{T.}~\bibnamefont{Kontos}},
  \bibinfo{journal}{Phys. Rev. Lett.} \textbf{\bibinfo{volume}{107}},
  \bibinfo{pages}{256804} (\bibinfo{year}{2011}).

\bibitem[{\citenamefont{Schiro and Le~Hur}(2014)}]{PhysRevB.89.195127}
\bibinfo{author}{\bibfnamefont{M.}~\bibnamefont{Schiro}} \bibnamefont{and}
  \bibinfo{author}{\bibfnamefont{K.}~\bibnamefont{Le~Hur}},
  \bibinfo{journal}{Phys. Rev. B} \textbf{\bibinfo{volume}{89}},
  \bibinfo{pages}{195127} (\bibinfo{year}{2014}).

\bibitem[{\citenamefont{Bergenfeldt and Samuelsson}(2012)}]{PhysRevB.85.045446}
\bibinfo{author}{\bibfnamefont{C.}~\bibnamefont{Bergenfeldt}} \bibnamefont{and}
  \bibinfo{author}{\bibfnamefont{P.}~\bibnamefont{Samuelsson}},
  \bibinfo{journal}{Phys. Rev. B} \textbf{\bibinfo{volume}{85}},
  \bibinfo{pages}{045446} (\bibinfo{year}{2012}).

\bibitem[{\citenamefont{Roy and Hughes}(2011)}]{PhysRevX.1.021009}
\bibinfo{author}{\bibfnamefont{C.}~\bibnamefont{Roy}} \bibnamefont{and}
  \bibinfo{author}{\bibfnamefont{S.}~\bibnamefont{Hughes}},
  \bibinfo{journal}{Phys. Rev. X} \textbf{\bibinfo{volume}{1}},
  \bibinfo{pages}{021009} (\bibinfo{year}{2011}).

\bibitem[{\citenamefont{Childress et~al.}(2004)\citenamefont{Childress,
  S{\o}rensen, and Lukin}}]{PhysRevA.69.042302}
\bibinfo{author}{\bibfnamefont{L.}~\bibnamefont{Childress}},
  \bibinfo{author}{\bibfnamefont{A.~S.} \bibnamefont{S{\o}rensen}},
  \bibnamefont{and} \bibinfo{author}{\bibfnamefont{M.~D.} \bibnamefont{Lukin}},
  \bibinfo{journal}{Phys. Rev. A} \textbf{\bibinfo{volume}{69}}
  (\bibinfo{year}{2004}).

\bibitem[{\citenamefont{Jin et~al.}(2011)\citenamefont{Jin, Marthaler, Cole,
  Shnirman, and Sch\"on}}]{1103.5051}
\bibinfo{author}{\bibfnamefont{P.-Q.} \bibnamefont{Jin}},
  \bibinfo{author}{\bibfnamefont{M.}~\bibnamefont{Marthaler}},
  \bibinfo{author}{\bibfnamefont{J.~H.} \bibnamefont{Cole}},
  \bibinfo{author}{\bibfnamefont{A.}~\bibnamefont{Shnirman}}, \bibnamefont{and}
  \bibinfo{author}{\bibfnamefont{G.}~\bibnamefont{Sch\"on}},
  \bibinfo{journal}{Phys. Rev. B} \textbf{\bibinfo{volume}{84}},
  \bibinfo{pages}{035322} (\bibinfo{year}{2011}).

\bibitem[{\citenamefont{Jin et~al.}(2013)\citenamefont{Jin, Marthaler, Jin,
  Golubev, and Sch\"on}}]{JinMarthalerLasing}
\bibinfo{author}{\bibfnamefont{J.}~\bibnamefont{Jin}},
  \bibinfo{author}{\bibfnamefont{M.}~\bibnamefont{Marthaler}},
  \bibinfo{author}{\bibfnamefont{P.-Q.} \bibnamefont{Jin}},
  \bibinfo{author}{\bibfnamefont{D.}~\bibnamefont{Golubev}}, \bibnamefont{and}
  \bibinfo{author}{\bibfnamefont{G.}~\bibnamefont{Sch\"on}},
  \bibinfo{journal}{New J. Phys.} \textbf{\bibinfo{volume}{15}},
  \bibinfo{pages}{025044} (\bibinfo{year}{2013}).

\bibitem[{\citenamefont{Vorojtsov et~al.}(2005)\citenamefont{Vorojtsov,
  Mucciolo, and Baranger}}]{PhysRevB.71.205322}
\bibinfo{author}{\bibfnamefont{S.}~\bibnamefont{Vorojtsov}},
  \bibinfo{author}{\bibfnamefont{E.~R.} \bibnamefont{Mucciolo}},
  \bibnamefont{and} \bibinfo{author}{\bibfnamefont{H.~U.}
  \bibnamefont{Baranger}}, \bibinfo{journal}{Phys. Rev. B}
  \textbf{\bibinfo{volume}{71}}, \bibinfo{pages}{205322}
  (\bibinfo{year}{2005}).

\bibitem[{\citenamefont{Gullans et~al.}(2015)\citenamefont{Gullans, Liu,
  Stehlik, Petta, and Taylor}}]{PhysRevLett.114.196802}
\bibinfo{author}{\bibfnamefont{M.~J.} \bibnamefont{Gullans}},
  \bibinfo{author}{\bibfnamefont{Y.-Y.} \bibnamefont{Liu}},
  \bibinfo{author}{\bibfnamefont{J.}~\bibnamefont{Stehlik}},
  \bibinfo{author}{\bibfnamefont{J.~R.} \bibnamefont{Petta}}, \bibnamefont{and}
  \bibinfo{author}{\bibfnamefont{J.~M.} \bibnamefont{Taylor}},
  \bibinfo{journal}{Phys. Rev. Lett.} \textbf{\bibinfo{volume}{114}},
  \bibinfo{pages}{196802} (\bibinfo{year}{2015}).

\bibitem[{\citenamefont{Marthaler et~al.}(2015)\citenamefont{Marthaler, Utsumi,
  and Golubev}}]{1503.01597}
\bibinfo{author}{\bibfnamefont{M.}~\bibnamefont{Marthaler}},
  \bibinfo{author}{\bibfnamefont{Y.}~\bibnamefont{Utsumi}}, \bibnamefont{and}
  \bibinfo{author}{\bibfnamefont{D.~S.} \bibnamefont{Golubev}},
  \bibinfo{journal}{Phys. Rev. B} \textbf{\bibinfo{volume}{91}},
  \bibinfo{pages}{184515} (\bibinfo{year}{2015}).

\bibitem[{\citenamefont{Frey et~al.}(2012)\citenamefont{Frey, Leek, Beck,
  Blais, Ihn, Ensslin, and Wallraff}}]{PhysRevLett.108.046807}
\bibinfo{author}{\bibfnamefont{T.}~\bibnamefont{Frey}},
  \bibinfo{author}{\bibfnamefont{P.~J.} \bibnamefont{Leek}},
  \bibinfo{author}{\bibfnamefont{M.}~\bibnamefont{Beck}},
  \bibinfo{author}{\bibfnamefont{A.}~\bibnamefont{Blais}},
  \bibinfo{author}{\bibfnamefont{T.}~\bibnamefont{Ihn}},
  \bibinfo{author}{\bibfnamefont{K.}~\bibnamefont{Ensslin}}, \bibnamefont{and}
  \bibinfo{author}{\bibfnamefont{A.}~\bibnamefont{Wallraff}},
  \bibinfo{journal}{Phys. Rev. Lett.} \textbf{\bibinfo{volume}{108}},
  \bibinfo{pages}{046807} (\bibinfo{year}{2012}).

\bibitem[{\citenamefont{Basset et~al.}(2013)\citenamefont{Basset, Jarausch,
  Stockklauser, Frey, Reichl, Wegscheider, Ihn, Ensslin, and
  Wallraff}}]{PhysRevB.88.125312}
\bibinfo{author}{\bibfnamefont{J.}~\bibnamefont{Basset}},
  \bibinfo{author}{\bibfnamefont{D.-D.} \bibnamefont{Jarausch}},
  \bibinfo{author}{\bibfnamefont{A.}~\bibnamefont{Stockklauser}},
  \bibinfo{author}{\bibfnamefont{T.}~\bibnamefont{Frey}},
  \bibinfo{author}{\bibfnamefont{C.}~\bibnamefont{Reichl}},
  \bibinfo{author}{\bibfnamefont{W.}~\bibnamefont{Wegscheider}},
  \bibinfo{author}{\bibfnamefont{T.~M.} \bibnamefont{Ihn}},
  \bibinfo{author}{\bibfnamefont{K.}~\bibnamefont{Ensslin}}, \bibnamefont{and}
  \bibinfo{author}{\bibfnamefont{A.}~\bibnamefont{Wallraff}},
  \bibinfo{journal}{Phys. Rev. B} \textbf{\bibinfo{volume}{88}},
  \bibinfo{pages}{125312} (\bibinfo{year}{2013}).

\bibitem[{\citenamefont{Liu et~al.}(2015)\citenamefont{Liu, Stehlik, Eichler,
  Gullans, Taylor, and Petta}}]{285.full}
\bibinfo{author}{\bibfnamefont{Y.-Y.} \bibnamefont{Liu}},
  \bibinfo{author}{\bibfnamefont{J.}~\bibnamefont{Stehlik}},
  \bibinfo{author}{\bibfnamefont{C.}~\bibnamefont{Eichler}},
  \bibinfo{author}{\bibfnamefont{M.~J.} \bibnamefont{Gullans}},
  \bibinfo{author}{\bibfnamefont{J.~M.} \bibnamefont{Taylor}},
  \bibnamefont{and} \bibinfo{author}{\bibfnamefont{J.~R.} \bibnamefont{Petta}},
  \bibinfo{journal}{Science} \textbf{\bibinfo{volume}{347}},
  \bibinfo{pages}{285} (\bibinfo{year}{2015}).

\bibitem[{\citenamefont{Liu et~al.}(2014)\citenamefont{Liu, Petersson, Stehlik,
  Taylor, and Petta}}]{PhysRevLett.113.036801}
\bibinfo{author}{\bibfnamefont{Y.-Y.} \bibnamefont{Liu}},
  \bibinfo{author}{\bibfnamefont{K.~D.} \bibnamefont{Petersson}},
  \bibinfo{author}{\bibfnamefont{J.}~\bibnamefont{Stehlik}},
  \bibinfo{author}{\bibfnamefont{J.~M.} \bibnamefont{Taylor}},
  \bibnamefont{and} \bibinfo{author}{\bibfnamefont{J.~R.} \bibnamefont{Petta}},
  \bibinfo{journal}{Phys. Rev. Lett.} \textbf{\bibinfo{volume}{113}},
  \bibinfo{pages}{036801} (\bibinfo{year}{2014}).

\bibitem[{\citenamefont{Stockklauser et~al.}(2015)\citenamefont{Stockklauser,
  Maisi, Basset, Cujia, Reichl, Wegscheider, Ihn, Wallraff, and
  Ensslin}}]{PhysRevLett.115.046802}
\bibinfo{author}{\bibfnamefont{A.}~\bibnamefont{Stockklauser}},
  \bibinfo{author}{\bibfnamefont{V.~F.} \bibnamefont{Maisi}},
  \bibinfo{author}{\bibfnamefont{J.}~\bibnamefont{Basset}},
  \bibinfo{author}{\bibfnamefont{K.}~\bibnamefont{Cujia}},
  \bibinfo{author}{\bibfnamefont{C.}~\bibnamefont{Reichl}},
  \bibinfo{author}{\bibfnamefont{W.}~\bibnamefont{Wegscheider}},
  \bibinfo{author}{\bibfnamefont{T.}~\bibnamefont{Ihn}},
  \bibinfo{author}{\bibfnamefont{A.}~\bibnamefont{Wallraff}}, \bibnamefont{and}
  \bibinfo{author}{\bibfnamefont{K.}~\bibnamefont{Ensslin}},
  \bibinfo{journal}{Phys. Rev. Lett.} \textbf{\bibinfo{volume}{115}},
  \bibinfo{pages}{046802} (\bibinfo{year}{2015}).

\bibitem[{\citenamefont{Colless et~al.}(2014)\citenamefont{Colless, Croot,
  Stace, Doherty, Barrett, Lu, Gossard, and Reilly}}]{ncomms4716}
\bibinfo{author}{\bibfnamefont{J.}~\bibnamefont{Colless}},
  \bibinfo{author}{\bibfnamefont{X.}~\bibnamefont{Croot}},
  \bibinfo{author}{\bibfnamefont{T.}~\bibnamefont{Stace}},
  \bibinfo{author}{\bibfnamefont{A.}~\bibnamefont{Doherty}},
  \bibinfo{author}{\bibfnamefont{S.}~\bibnamefont{Barrett}},
  \bibinfo{author}{\bibfnamefont{H.}~\bibnamefont{Lu}},
  \bibinfo{author}{\bibfnamefont{A.}~\bibnamefont{Gossard}}, \bibnamefont{and}
  \bibinfo{author}{\bibfnamefont{D.}~\bibnamefont{Reilly}},
  \bibinfo{journal}{Nat. Commun.} \textbf{\bibinfo{volume}{5}},
  \bibinfo{pages}{1} (\bibinfo{year}{2014}).

\bibitem[{\citenamefont{Viennot et~al.}(2014)\citenamefont{Viennot, Delbecq,
  Dartiailh, Cottet, and Kontos}}]{PhysRevB.89.165404}
\bibinfo{author}{\bibfnamefont{J.~J.} \bibnamefont{Viennot}},
  \bibinfo{author}{\bibfnamefont{M.~R.} \bibnamefont{Delbecq}},
  \bibinfo{author}{\bibfnamefont{M.~C.} \bibnamefont{Dartiailh}},
  \bibinfo{author}{\bibfnamefont{A.}~\bibnamefont{Cottet}}, \bibnamefont{and}
  \bibinfo{author}{\bibfnamefont{T.}~\bibnamefont{Kontos}},
  \bibinfo{journal}{Phys. Rev. B} \textbf{\bibinfo{volume}{89}}
  (\bibinfo{year}{2014}).

\bibitem[{\citenamefont{Karlewski and Marthaler}(2014)}]{PhysRevB.90.104302}
\bibinfo{author}{\bibfnamefont{C.}~\bibnamefont{Karlewski}} \bibnamefont{and}
  \bibinfo{author}{\bibfnamefont{M.}~\bibnamefont{Marthaler}},
  \bibinfo{journal}{Phys. Rev. B} \textbf{\bibinfo{volume}{90}},
  \bibinfo{pages}{104302} (\bibinfo{year}{2014}).

\bibitem[{\citenamefont{Vogt et~al.}(2012)\citenamefont{Vogt, Cole, Marthaler,
  and Sch\"on}}]{PhysRevB.85.174515}
\bibinfo{author}{\bibfnamefont{N.}~\bibnamefont{Vogt}},
  \bibinfo{author}{\bibfnamefont{J.~H.} \bibnamefont{Cole}},
  \bibinfo{author}{\bibfnamefont{M.}~\bibnamefont{Marthaler}},
  \bibnamefont{and} \bibinfo{author}{\bibfnamefont{G.}~\bibnamefont{Sch\"on}},
  \bibinfo{journal}{Phys. Rev. B} \textbf{\bibinfo{volume}{85}},
  \bibinfo{pages}{174515} (\bibinfo{year}{2012}).

\bibitem[{\citenamefont{Fano}(1957)}]{RevModPhys.29.74}
\bibinfo{author}{\bibfnamefont{U.}~\bibnamefont{Fano}}, \bibinfo{journal}{Rev.
  Mod. Phys.} \textbf{\bibinfo{volume}{29}}, \bibinfo{pages}{74}
  (\bibinfo{year}{1957}).

\bibitem[{\citenamefont{Brandes}(2005)}]{1-s2.0-S0370157304005496-main}
\bibinfo{author}{\bibfnamefont{T.}~\bibnamefont{Brandes}},
  \bibinfo{journal}{Phys. Rep.} \textbf{\bibinfo{volume}{408}},
  \bibinfo{pages}{315} (\bibinfo{year}{2005}).

\bibitem[{\citenamefont{Tawara et~al.}(2008)\citenamefont{Tawara, Kamada,
  Zhang, Tanabe, Cade, Ding, Johnson, Gotoh, Kuramochi, Notomi
  et~al.}}]{oe-16-8-5199}
\bibinfo{author}{\bibfnamefont{T.}~\bibnamefont{Tawara}},
  \bibinfo{author}{\bibfnamefont{H.}~\bibnamefont{Kamada}},
  \bibinfo{author}{\bibfnamefont{Y.-H.} \bibnamefont{Zhang}},
  \bibinfo{author}{\bibfnamefont{T.}~\bibnamefont{Tanabe}},
  \bibinfo{author}{\bibfnamefont{N.~I.} \bibnamefont{Cade}},
  \bibinfo{author}{\bibfnamefont{D.}~\bibnamefont{Ding}},
  \bibinfo{author}{\bibfnamefont{S.~R.} \bibnamefont{Johnson}},
  \bibinfo{author}{\bibfnamefont{H.}~\bibnamefont{Gotoh}},
  \bibinfo{author}{\bibfnamefont{E.}~\bibnamefont{Kuramochi}},
  \bibinfo{author}{\bibfnamefont{M.}~\bibnamefont{Notomi}},
  \bibnamefont{et~al.}, \bibinfo{journal}{Opt. Exp.}
  \textbf{\bibinfo{volume}{16}}, \bibinfo{pages}{5199} (\bibinfo{year}{2008}).

\bibitem[{\citenamefont{Megrant et~al.}(2012)\citenamefont{Megrant, Neill,
  Barends, Chiaro, Chen, Feigl, Kelly, Lucero, Mariantoni, O’Malley
  et~al.}}]{Megrant2012}
\bibinfo{author}{\bibfnamefont{A.}~\bibnamefont{Megrant}},
  \bibinfo{author}{\bibfnamefont{C.}~\bibnamefont{Neill}},
  \bibinfo{author}{\bibfnamefont{R.}~\bibnamefont{Barends}},
  \bibinfo{author}{\bibfnamefont{B.}~\bibnamefont{Chiaro}},
  \bibinfo{author}{\bibfnamefont{Y.}~\bibnamefont{Chen}},
  \bibinfo{author}{\bibfnamefont{L.}~\bibnamefont{Feigl}},
  \bibinfo{author}{\bibfnamefont{J.}~\bibnamefont{Kelly}},
  \bibinfo{author}{\bibfnamefont{E.}~\bibnamefont{Lucero}},
  \bibinfo{author}{\bibfnamefont{M.}~\bibnamefont{Mariantoni}},
  \bibinfo{author}{\bibfnamefont{P.~J.~J.} \bibnamefont{O’Malley}},
  \bibnamefont{et~al.}, \bibinfo{journal}{App. Phys. Lett.}
  \textbf{\bibinfo{volume}{100}}, \bibinfo{pages}{113510}
  (\bibinfo{year}{2012}).

\bibitem[{\citenamefont{Petersson et~al.}(2012)\citenamefont{Petersson, McFaul,
  Schroer, Jung, Taylor, Houck, and Petta}}]{nature11559}
\bibinfo{author}{\bibfnamefont{K.~D.} \bibnamefont{Petersson}},
  \bibinfo{author}{\bibfnamefont{L.~W.} \bibnamefont{McFaul}},
  \bibinfo{author}{\bibfnamefont{M.~D.} \bibnamefont{Schroer}},
  \bibinfo{author}{\bibfnamefont{M.}~\bibnamefont{Jung}},
  \bibinfo{author}{\bibfnamefont{J.~M.} \bibnamefont{Taylor}},
  \bibinfo{author}{\bibfnamefont{A.~A.} \bibnamefont{Houck}}, \bibnamefont{and}
  \bibinfo{author}{\bibfnamefont{J.~R.} \bibnamefont{Petta}},
  \bibinfo{journal}{Nature} \textbf{\bibinfo{volume}{490}},
  \bibinfo{pages}{380–383} (\bibinfo{year}{2012}).

\bibitem[{\citenamefont{Viennot et~al.}(2015)\citenamefont{Viennot, Dartiailh,
  Cottet, and Kontos}}]{Science-2015-Viennot-408-11}
\bibinfo{author}{\bibfnamefont{J.~J.} \bibnamefont{Viennot}},
  \bibinfo{author}{\bibfnamefont{M.~C.} \bibnamefont{Dartiailh}},
  \bibinfo{author}{\bibfnamefont{A.}~\bibnamefont{Cottet}}, \bibnamefont{and}
  \bibinfo{author}{\bibfnamefont{T.}~\bibnamefont{Kontos}},
  \bibinfo{journal}{Science} \textbf{\bibinfo{volume}{349}},
  \bibinfo{pages}{408} (\bibinfo{year}{2015}).

\end{thebibliography}

\end{document}